\documentclass[12pt]{iopart}
\usepackage{iopams}  
\usepackage{graphicx}
\usepackage{epstopdf}
\usepackage{cite}

\newcommand{\sign}{\mathrm{sign}}

\begin{document}

\title{A network model for field and quenched disorder effects in artificial spin ice.}

\author{Zoe~Budrikis$^{1, 2, 3}$, Paolo Politi$^{2, 4}$ and R L Stamps$^3$}
\ead{zoe.budrikis@gmail.com}
\address{$^1$School of Physics, The University of Western Australia, 35 Stirling Hwy, Crawley 6009, Australia}
\address{$^2$Istituto dei Sistemi Complessi CNR, Via Madonna del Piano 10, 50019 Sesto Fiorentino, Italy}
\address{$^3$SUPA School of Physics and Astronomy, University of Glasgow, Glasgow G12 8QQ, United Kingdom}
\address{$^4$INFN Sezione di Firenze, via G. Sansone 1, 50019 Sesto Fiorentino, Italy}

\begin{abstract}
We have performed a systematic study of the effects of field strength
and quenched disorder on the driven dynamics of square artificial spin ice.
We construct a network representation of the configurational phase space,
where nodes represent the microscopic configurations and a directed link
between node $i$ and node $j$ means that the field may induce a transition
between the corresponding configurations. In this way, we are able to
quantitatively describe how the field and the disorder affect the connectedness
of states and the reversibility of dynamics. In particular, we have shown that for
optimal field strengths, a substantial fraction of all states can be accessed
using external driving fields, and this fraction is increased by disorder. 
We discuss how this relates to control and potential information storage applications for artificial spin ices.
\end{abstract}

\submitto{\NJP}

\maketitle

\section{Introduction}
In an idealised model, the nanoscale magnetic islands of artificial spin ice~\cite{Wang:2006, Tanaka:2006, Qi:2008} are identical Ising macrospins. However, in reality, unavoidable small variations during the island fabrication process lead to a distribution of island properties. In fact, a comparison of experimental and simulation results for square artificial spin ice shows that the effective distribution of island magnetisation switching barriers has a width on the same scale as nearest-neighbour interactions \cite{Budrikis2011ConstH}. Quenched disorder provides pinning and nucleation sites and modifies hysteresis \cite{Libal:2009, Kohli2011, Ladak:2010, Mengotti:2010, Daunheimer2011, Libal2011a}. Accordingly, a complete understanding of artificial spin ice requires not only an understanding of the behaviour of frustrated coupled Ising spins subject to an external magnetic field, but also an understanding of the role of quenched disorder in those dynamics.

One approach to studying square artificial spin ice dynamics that has received much attention treats the vertices of the array as objects \cite{Wang:2006, Moller:2006, Libal:2006, Nisoli:2007, Remhof:2008, Ke:2008, Libal:2009, Mol:2009, Budrikis:2010, Nisoli:2010, Phatak2011, Morgan2011a, Budrikis2011ConstH, Budrikis2011disorder}. For example, the populations of different vertex types (distinguished by their energy) provide a measure of the level of ordering and can be analysed in terms of an effective temperature associated with ac demagnetisation; magnetisation reversal of an array can be characterised by the motion and interactions of `monopole' vertices; the effects of quenched disorder that affects interactions can be described in terms of variations in vertex energies; and the evolution under a rotating applied field can be modelled in terms of population dynamics of vertex types, allowing analytical expressions for the system's evolution to be written down and solved.

We have recently demonstrated~\cite{Budrikis2011networks} the value of an alternative approach, in which the fundamental objects of interest are not vertices but whole-array spin configurations. In this approach, the set of all Ising spin configurations forms a discrete phase space, and the action of an applied field is to `transport' the system from one point in phase space to another, via one or more spin flips. This picture is essentially a mapping of dynamics onto a directed network, in which configurational states are nodes and a directed link exists from node $i$ to node $f$, i.e., $i \to f$, if an applied field can drive the system from configuration $i$ to configuration $f$. In other words, the network describes which barriers to flipping spins may be overcome  by an external field. Related approaches involving network analysis have been used previously in the study of dynamical maps~\cite{Borges2007, Kyriakopoulos2007}, geometrically frustrated systems~\cite{Han:2009, Han:2010, Peng2011}, the random field Ising model~\cite{Bertotti:2007, Bortolotti:2008, Bortolotti:2010}, proteins~\cite{Rao2004, Noe2006, Gfeller2007a, Gfeller2007, Ravasz2007}, polymers~\cite{Scala2001}, atomic clusters~\cite{Doye2002} and glasses~\cite{Angelani1998, Angelani2000, Burda2007, Seyed-Allaei2008, Carmi2009, Baronchelli2009}.

In this work, we extend our previous results~\cite{Budrikis2011networks}, which showed that quenched disorder lifts degeneracies in how the magnetic moments respond to a global driving field, allowing access to states that cannot be accessed in a perfect system. Here, we study how not only disorder but also driving field strength affects the accessibility of states and controllability of dynamics.

The structure of this paper is as follows. In section~\ref{methods} we describe our model system and outline the methods we use. 
In section~\ref{degrees} we discuss how the degree of a network node, that is, the number of links pointing into or out of it, relates to the energy of the configuration it represents. 
In section~\ref{accessibility} we study the number of states that can be reached from a polarised configuration and the reversibility of dynamical transitions between those states. 
These results are built on in section~\ref{structure}, where we study the structure of the spin ice dynamics networks and discuss how small changes to the properties of the artificial spin ice can lead to large changes in dynamics. 
We give definitions throughout this paper of the network theoretic terminology and concepts used, but readers wishing for a more thorough introduction to network theory should see, for example, \cite{Albert:2002, Newman:2003, Boccaletti2006, Dorogovtsev:2010}.

\section{Methods}
\label{methods}

\subsection{Energetics, dynamics and disorder}
\label{energetics}

\begin{figure}
  \centering
  \includegraphics[width=0.7\columnwidth]{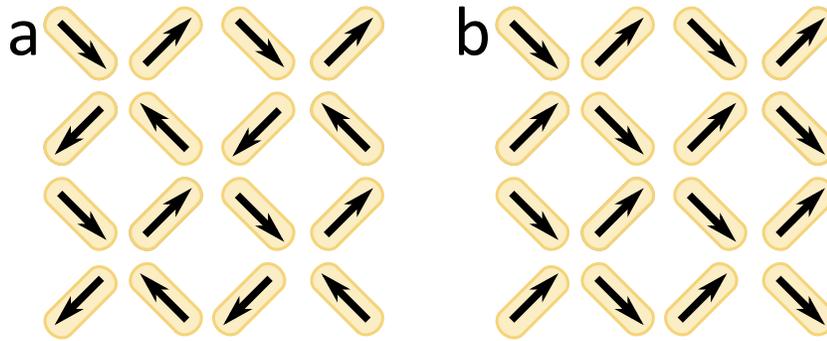}
  \caption{The geometry of the $4\times4$ array studied. (a) One of the two possible ground state tilings. The other ground state is obtained by flipping all spins. (b) One of the four possible polarised configurations, as would be obtained by applying a strong field in the $+x$ direction.}
  \label{array_geometry}
\end{figure}

In this work, we study a $4\times4$ spin ice array, with geometry shown in figure~\ref{array_geometry}. The 16 Ising spins of the array can take a total of $2^{16}=65,536$ configurations. We show elsewhere~\cite{Budrikis2011networks} that this small, feasible to analyse, system has dynamics that are sufficiently similar to those of larger arrays that the results presented here are broadly relevant, even if small arrays are limited in their support of configurational features such as domain walls that are seen in larger systems. Larger arrays are inaccessible to our network analysis due to exponential growth in the number of configurational states (network nodes) with system size. For example,  increasing the array size to even $5\times5$ spins increases the number of configurations -- or equivalently, network nodes -- to around $33,000,000$. Of course, numerical simulations of dynamics of larger systems are effectively a sampling of their phase space networks, but we leave a rigorous interpretation of simulated dynamics in terms of networks for future work.

We first describe our model for the energetics and dynamics of this system, and then describe how we construct a network representation of those dynamics.
The island magnetic moments are Ising point dipoles, interacting so that the dipolar energy of island $i$ is
\begin{eqnarray}
E_{\mathrm{dip}}^{(i)} 
	&=& -\mu_0 \vec h_{\mathrm{dip}}^{(i)}\cdot \vec{M}_i \\
	&=& \frac{\mu_0}{4 \pi} \sum_{j\ne i} \biggl(
	\frac{\vec{M_i} \cdot \vec{M_j}}{r_{ij}^3}
	- 3 \frac{(\vec{M_i} \cdot \vec{r_{ij}}) (\vec{M_i} \cdot \vec{r_{ij}})}{r_{ij}^5}
	\biggr).
\end{eqnarray}
We set both the island magnetic moment $M$ and the nearest-neighbour distance to unity, so that the nearest-neighbour coupling has strength $1.5$, in units of $\mu_0/(4\pi)$. This gives the energy scale of the island-island interactions. The total dipolar energy is given by $(1/2)\sum_i E_{\mathrm{dip}}^{(i)}$. A ground state spin configuration that minimises the total dipolar energy is shown in figure~\ref{array_geometry}(a); the other ground state configuration is obtained by a global spin flip. 

The other field acting on each island is the external field $\vec{h}$, which gives a Zeeman contribution to the total energy of spin $i$
\begin{equation}
E_{Z}^{(i)} = -\mu_0\vec{h}\cdot \vec{M}_i.
\end{equation}
When a field strong enough to overcome the dipolar interactions is applied at approximately $45^{\circ}$ to the island axes, the system's favoured state is a polarised configuration, in which all spins have the same projection onto the field. One of the four possible polarised configurations is shown in figure~\ref{array_geometry}(b). Polarised configurations are of particular interest because they are experimentally reproducible. 

In addition to the island interactions, a second energy scale in the system is a barrier to island switching. We model this using the switching criterion
\begin{equation}
\label{switching}
-(\vec{h}_{\mathrm{dip}}^{(i)} + \vec{h}) \cdot \hat{M}_i > h_c^{(i)},
\end{equation}
where $\hat{M}$ is the dimensionless unit vector along $\vec{M}$. In other words, we require that the component of the total field antiparallel to an island's magnetisation be greater than the island's intrinsic switching field $h_c^{(i)}$. A similar threshold-based model for switching has been used by other authors~\cite{Moller:2006, Remhof:2008, Mengotti:2010, Ladak:2010} . (We have studied other switching criteria, such as Stoner-Wohlfarth switching, in numerical simulations and find qualitatively similar dynamics.) In experimental systems, the island switching fields are usually designed to be larger than the dipolar fields so that configuration states can only change under external fields~\cite{Wang:2006}. Here we set the mean switching field to $h_c=11.25$, a value outside the range of dipolar coupling strengths.

The dynamics of the system under fixed external field $\vec{h}$ consists of a series of single spin flips, determined by criterion (\ref{switching}). The set of all spins that satisfy (\ref{switching}) is calculated, then one is chosen uniformly at random and flipped. The set of spins satisfying (\ref{switching}) is re-calculated for the new configuration, and again one is flipped at random. This process continues until no further spin flips are possible, at which point a `final', stationary configuration has been attained. Because spin flips are selected at random and the set of spins that can flip is recalculated at each step, more than one series of spin flips may be possible under application of the same field. In previous simulation studies~\cite{Budrikis:2010}, this meant that different simulation runs did not necessarily have the same outcome, and we instead averaged over outcomes. In our network studies, we enumerate over all these possibilities, as described in section~\ref{network_construction}.

As seen in equation~(\ref{switching}), the response of an island to the external driving field is controlled both by its interactions with other islands, and its intrinsic switching field, unless the external field is strong enough to overcome these. Inter-island interactions bias the response of the system towards low-energy states. At the same time, disorder, in the form of a distribution of island properties, introduces randomness in the response to fields. In this work, we focus exclusively on switching field disorder. We show elsewhere~\cite{Budrikis2011disorder} that other types of disorder, such as disorder in island positions or orientations, have a similar effect on dynamics as switching field disorder does, and that all disorder can be characterised in terms of an effective switching field disorder. Based on that result, in this work we consider only switching field disorder. A direction for future study might be a study of disorder with correlations.

We characterise the strength of disorder by $\sigma$, the standard deviation of the switching fields for all islands. The switching fields are drawn from a nominally Gaussian distribution. However, because in each disorder realisation only 16 $h_c^{(i)}$ values are required, the mean of the generated pseudo-random numbers can deviate substantially from the nominal mean of $11.25$. This change in mean switching field can have a significant effect on dynamics, so we `correct' each $h_c^{(i)}$ value by $-\Delta$, where $\Delta$ is the difference between the mean of the generated values and $11.25$. Thus, for each disorder realisation studied, the mean switching field is exactly $11.25$, the same as the switching field used in the absence of disorder, allowing meaningful comparison between disorder realisations.

In our studies on field strength, we  analyse three networks at each field strength: one for the perfect system, and two realisations of disorder. Both realisations are in the strong disorder regime of \cite{Budrikis2011disorder}, with $\sigma=2.05$ and $\sigma=2.22$ both larger than the scale of dipolar interactions. We focus on two similar disorder strengths in order to verify that networks representing systems with similar disorder strength have similar properties, regardless of field strength.

\subsection{Network construction}
\label{network_construction}
The $4\times4$ Ising spin system we study has $2^{16}=65,536$ possible microscopic configurations. As mentioned in the Introduction, each configuration is a network node, and the number of nodes is fixed at $2^{16}$ for all networks we study. For a given external field $\vec{h}$, a link exists from node $i$ to node $f$ if the configuration corresponding to $i$ can evolve into the configuration corresponding to $f$ under $\vec{h}$, via a cascade of spin flips according to the dynamics described above. As we describe below, the network can be represented by a $2^{16}\times2^{16}$ matrix whose rows and columns represent nodes, and whose entries represent links (an `adjacency matrix', in the language of network theory).

The links of our networks are directed, that is, the existence of a link $i \to f$  does not necessarily imply the existence of a link $f \to i$. The reason for this is that dynamical transitions are not, in general, reversible: the athermal field-driven dynamics must involve transitions that lower the sum of dipolar and Zeeman energies, due to the nature of the energy barriers in the system. Directed networks appear in many other contexts, such as the world wide web \cite{Broder2000},  networks of corporate ownership and control \cite{Vitali2011}, and the network representation of basins of mutually-reachable states in the random field Ising model \cite{Bertotti:2007, Bortolotti:2008, Bortolotti:2010}. In contrast, the networks used to describe, e.g, the six-vertex model \cite{Han:2009, Han:2010, Peng2011} are undirected, because dynamics in those systems are taken to be reversible at the microscopic level.

The set links of the network depend on the fields used to construct the network. For simplicity, in this work, each network we consider is constructed for a single field amplitude $h$ and for field angles $\theta=0,\pi/128, 2\pi/128,\ldots$. This choice of field angles gives a network with properties approaching the expected finite limit for a continuously varying field angle, while remaining computationally tractable \cite{Budrikis2011networks}. The network also depends on the disorder realisation used, with different realisations giving different networks. 

In order to enumerate all network links, we first determine all allowed single spin flips for each of the $2^{16}$ configurations, for all fields $(h, \{\theta\})$. These can be stored as transition matrices $T(h, \theta)$, where $T_{ij}(h, \theta)=1$ if configuration $i$ can be transformed into configuration $j$ by a single spin flip that is allowed under a field $(h, \theta)$ according to criterion (\ref{switching}), and $T_{ij}(h, \theta)=0$ otherwise. Note that $T(h, \theta)$ is not symmetric, because only flips that lower the system's total (dipolar and Zeeman) energy are allowed.

The non-zero entries of $T^2$ give permitted transitions involving two spin flips. Similarly, three spin flips are described by $T^3$, and so on. The non-zero entries of $T^i$ are not all equal to $1$, so we apply the $\sign$ function ($\sign(x)=1$ if $x>0$ and 0 otherwise) to each element of $T^i$ to `normalise' the matrix.
Because there are 16 spins in the system and spins can only flip once in a dynamical cascade, the maximum length of a sequence of spin flips is 16, and we must have $\sign(T^{i})=\sign(T^{16})$ for all $i>16$. We denote $\sign(T^{16})$ as $A(h, \theta)$, and $A_{ij}(h, \theta)=1$ if configuration $i$ can evolve by a cascade of spin flips into configuration $j$, under the field $(h, \theta)$, and $A_{ij}(h, \theta)=0$ otherwise. We define the network adjacency matrix $\tilde{A}(h)$ by $\tilde{A}(h)=\sign(\sum_{\theta}A(h, \theta))$, where the sum is over all $\theta$ between $0$ and $2\pi$. $\tilde{A}$ describes the network representation of all possible dynamics under fields $(h, \{\theta\})$, and it is the properties of $\tilde{A}$ for various field strengths and disorder realisations that will concern us in the rest of this paper. 
We emphasise that this method of network construction is an exact enumeration over all possible transitions between configuration states allowed for a given set of external fields. 

We also emphasise that each network link is active only for a range of field angles $\theta_{\rm min} < \theta < \theta_{\rm max}$. Accordingly, the existence of a network path from a node $i$ to a node $j$ indicates that there exists one or more sequences of field angles that can be applied to a system prepared in state $i$ to drive it into state $j$. A particular sequence of field angles, also known as a `field protocol' in the literature, corresponds to the paths on the network that are generated by following links that are active for each angle in the sequence in turn. Conversely, the network contains information about all possible field protocols.

\section{Node degrees and energetics}
\label{degrees}
In this section we discuss how an entirely local property of network nodes, namely their degree, can be related to spin ice physics. We also discuss briefly what the distribution of node degrees shows about the global topology of the network; however, we will see in subsequent sections that the degree distributions are insufficient to completely describe the network topology.

The in- (out-) degree of a network node is the number of links pointing into (out of) it. In an undirected network, the two quantities are the same, but in a directed network they are different. The degree distribution ${\cal N}(k_{\mathrm{in (out)}})$ is the distribution of the number of nodes with degree $k_{\mathrm{in (out)}}$. The in degree of node $v$ is given by
\begin{equation}
k_{\rm in} = \sum_i \tilde{A}_{iv} - \tilde{A}_{vv},
\end{equation} 
and the out degree is given by 
\begin{equation}
k_{\rm out} = \sum_j \tilde{A}_{vj} - \tilde{A}_{vv}.
\end{equation}

\begin{figure}
  \centering
  \includegraphics[width=\columnwidth]{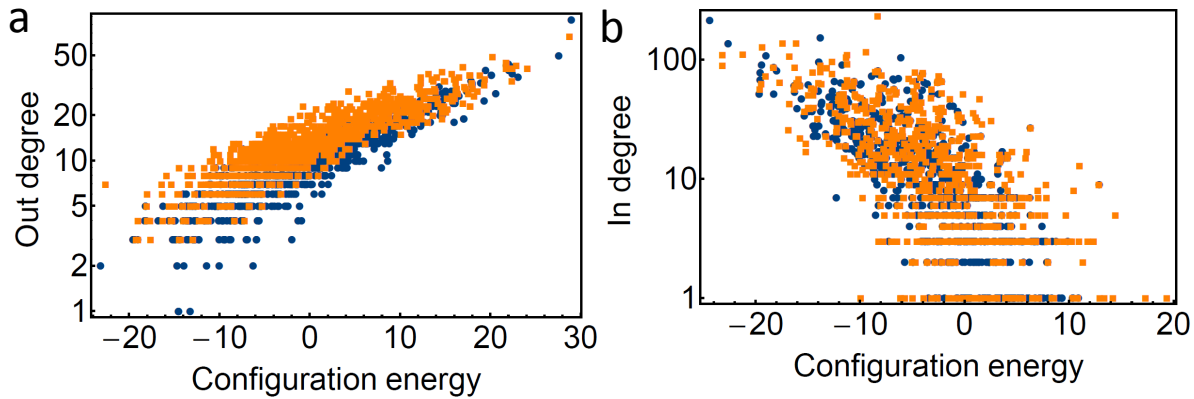}
  \caption{The (a) out degree and (b) in degree of 1000 randomly selected network nodes (spin configurations), plotted against their dipolar energy. Blue (dark grey) circles represent the perfect system, orange (light grey) squares represent the disordered system with $\sigma=2.05$.}
  \label{degree_vs_energy_perfect_dis}
\end{figure}

Figure~\ref{degree_vs_energy_perfect_dis} shows the in- and out-degrees of 1000 randomly selected nodes, \textit{vs} the energy of the spin configurations they represent, for networks describing a perfect and a disordered system (with $\sigma=2.05$) subject to a field of amplitude $h=11.5$. For both perfect and disordered systems, high-energy configurations correspond to nodes with low in-degree and high out-degree, and low-energy configurations correspond to nodes with a higher in-degree and lower out-degree.

Physically, a node with high out-degree represents a configuration that can `decay' into many others when fields are applied, while a node with low out-degree represents a `stable' configuration. 
`Stability' here refers to how a configuration can be modified by an external field, as measured by, for example, the probability that a field with direction chosen uniformly at random is able to drive the system to a new configuration.
Although the stability of a configuration is not completely correlated with its energy and the out-degree \textit{vs} energy data displays some spread, there is a clear relationship between energy and out-degree, as seen in figure~\ref{degree_vs_energy_perfect_dis}(a). This can be understood by considering the extremes of high and low energy: At the field strength studied here ($h=11.5$), all spins in a configuration that maximises dipolar energy can be flipped by an external field even when moderate disorder is present, so there are many pathways out of that configuration. On the other hand, the ground state of the system is stable even under moderate disorder. 

The energy dependence of the in-degree has a wider spread, especially when
in-degree is low.
This is because the barriers to `entering' a state are topological as well as energetic. 
A low in-degree may correspond to a state with high energy or it may correspond to a low energy state with antiferromagnetic ordering that is `hard' to access with a global external field, which tends to create ferromagnetic ordering. We discuss these ideas in relation to the ground state of the system elsewhere \cite{Budrikis2011ConstH}. Nevertheless, it is the case that only low-energy configurations have high in-degree, as seen in figure~\ref{degree_vs_energy_perfect_dis}(b).

Node degrees can also give information about the global network topology, via the degree distribution ${\cal N}(k)$. For example, many real-world networks such as the world wide web and networks of scientific citation display power-law degree distributions \cite{Albert:2002, Newman:2003}. On the other hand, in Erd\"{o}s-R\'{e}nyi random graphs \cite{Erdos1959}, where an undirected link between any pair of nodes is present with probability $p$ and absent with probability $1-p$, the degree distribution is Poissonian. Networks describing the phase space of other frustrated spin models and lattice gas models have been shown to have Gaussian degree distributions \cite{Han:2009, Han:2010, Peng2011}.

\begin{figure}
  \centering
  \includegraphics[width=\columnwidth]{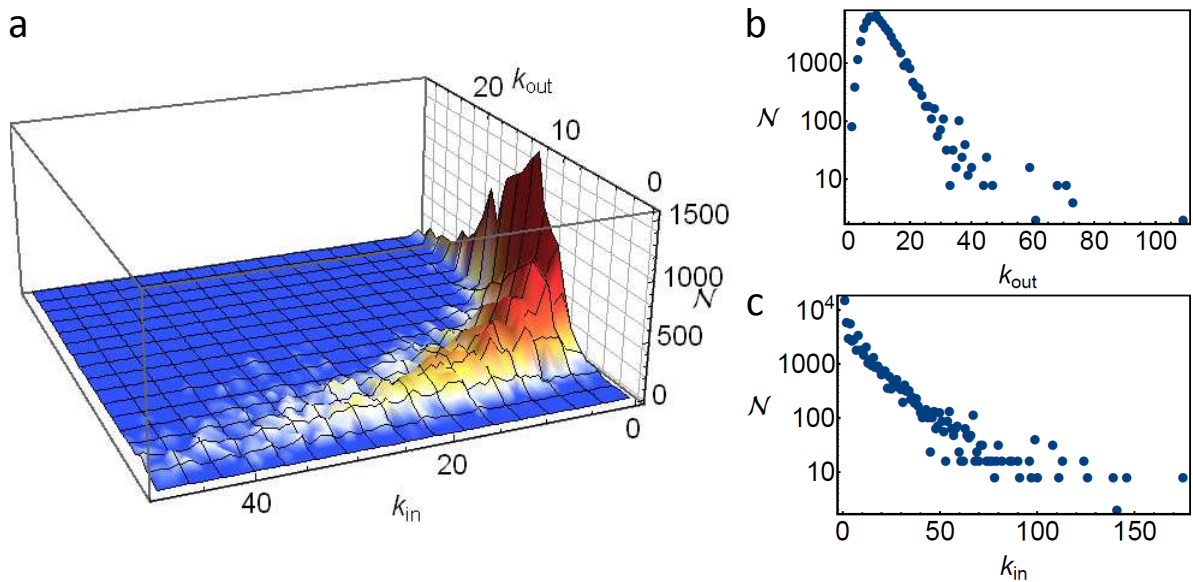}
  \caption{Degree distributions for a perfect system, subject to a field $h=11$. (a) The joint in- and out-degree distribution, for $k_{\rm in}<50$ and $k_{\rm out}<30$. (b) The separate out-degree distribution. (c) The separate in-degree distribution. Disordered systems have similar degree distributions.}
  \label{degree_distributions}
\end{figure}

Directed networks can be described by three distributions: the joint in- and out-degree distribution, which gives the probability that a randomly selected node has in-degree $k_{\rm in}$ and out-degree $k_{\rm out}$, and the two separate degree distributions, which are obtained by integrating the joint distribution. In figure~\ref{degree_distributions}, we plot these three distributions for the network describing an undisordered spin ice at $h=11$. In contrast to other frustrated systems \cite{Han:2009, Han:2010, Peng2011}, the spin ice networks do not have a Gaussian degree distribution. Instead, the distributions are clearly asymmetric, and the separate distributions for in- and out-degree are different. As expected from our discussion above about the relationship between configuration energy and node degree, there is a tendency for low in-degree nodes to have a higher out-degree, and vice-versa. However, we will see in section~\ref{structure} that the degree distribution is inadequate to completely characterise the network.

\section{Accessibility of states and reversibility of dynamics}
\label{accessibility}
Many potential technological applications of artificial spin ices and related systems depend on the answers to two questions. How many distinct configurational states are available, and can those states be accessed reliably? These questions are particularly important for applications relating to information storage: although a system of $N$ Ising spins has in principle $2^N$ configurations and can store $N$ bits of information, the effective information capacity is lower if only some fraction of those configurations can be realised. In this section, we see that network tools are ideal for studying these questions, and for uncovering effects of disorder and field strength on the accessibility of states.

As already mentioned in section~\ref{energetics}, the only magnetisation configurations currently known to be exactly experimentally reproducible are the four polarised configurations (see figure~\ref{array_geometry}(b)). Because these configurations can be reliably obtained, in almost all experimental and simulation studies to date, the initial state of the system has been polarised \cite{Wang:2006, Libal:2006, Remhof:2008, Ke:2008, Libal:2009, Budrikis:2010, Phatak2011, Morgan2011a, Libal2011a, Kapaklis2011}. 
It is therefore of particular interest to study the accessibility of states from these configurations.

We focus on states that can be accessed using a sequence of fields with fixed amplitude. As mentioned in section~\ref{network_construction}, such a sequence of fields corresponds to a `walk' on the network, in which at each step only a link that is active at that field angle is followed. There may be more than one walk corresponding to a sequence of fields, since there may be more than one link active for a given field angle. A random field protocol, in which the field angle is selected uniformly at random from $[0, 2\pi)$ at each step of the sequence, is essentially a random walk on the network~\footnote{The equivalence between the simulation of a random field protocol and a random walk on the network is not exact, because the network is unweighted and the probability a link is followed in a random walk depends only on the number of links out of a node, not the range of field angles over which they are active.}.

The total set of nodes that can be reached from an initial node $v$, following network paths of any length, is given by the fixed point of repeatedly multiplying the unit vector $\vec{v}$ (all entries zero, except for non-zero entry $v$) by the adjacency matrix $\tilde{A}$. Each state can be accessed via one or more field protocols, but a given field protocol may not be able to access all of them. We return to this question of ergodicity below.

\begin{figure}
  \centering
  \includegraphics[width=\columnwidth]{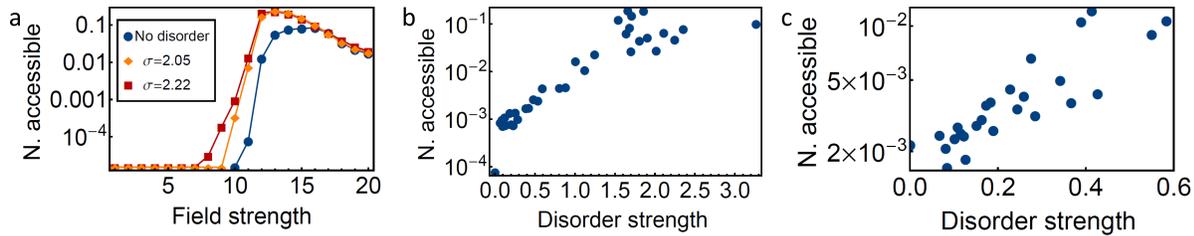}
  \caption{(a) The fraction of configurations that can be reached from an initial polarised configuration, \textit{vs} applied field amplitude. (b, c) The fraction of configurations that can be reached from an initial polarised configuration, \textit{vs} disorder strength (standard deviation in island switching fields), for an applied field of strength $h=11.5$ (b) and $h=11.75$ (c). For $h=11.5$, there is a jump in the number of accessible states when disorder is turned on, but for $h=11.75$, the number of accessible states increases smoothly. Each point represents a single disorder realisation. In all figures, the initial configuration is counted in the number of accessible states.}
  \label{n_reachable_polarised}
\end{figure}

Figure~\ref{n_reachable_polarised}(a) shows how the fraction of configurations that can be reached from the $+x$ polarised configuration depends on field strength, for both the perfect system and two disorder realisations ($\sigma=2.05$ and $\sigma=2.22$). For weak fields, the $+x$ polarised configuration is stable against applied field and no other states are accessible from it. However, for optimal field values, approximately $10\%$ of all configurations can be reached from the $+x$ polarised configuration. In the very high field limit (not shown), applied fields always polarise the system, so from the $+x$ polarised configuration there are 3 other states accessible, namely, the other polarised states.

Although the curves for the perfect and disordered systems in figure~\ref{n_reachable_polarised}(a) have similar form, for applied field strengths near the mean island switching field ($h_c=11.25$) the difference between perfect and disordered systems is large. This is in agreement with previous results~\cite{Budrikis2011networks}. Figure~\ref{n_reachable_polarised}(b) shows how the number of states accessible from the $+x$ polarised configuration depends on disorder strength, for $h=11.5$. There is a jump in the number of accessible states when disorder is turned on. This is because two of the configurations that can be reached from the $+x$ polarised configuration have spins that, in the perfect system, require an external field of $11.74$ to switch. A small disorder-induced decrease in the switching barrier for these spins allows them to flip at $h=11.5$, opening new dynamical pathways. This interpretation is confirmed by the network for $h=11.75$. For that field strength, the fraction of states accessible from the $+x$ polarised configuration  in the undisordered system is $\sim10^{-3}$, and increases smoothly with disorder strength, as shown in Figure~\ref{n_reachable_polarised}(c). The large impact of small disorder-induced changes to the system is a recurring theme in this work.

The number of states that can be reached from an initial polarised state is a starting point for describing dynamics, but this quantity does not give a complete picture. One question it gives little information about is that of ergodicity -- once a transition has been made from the polarised state to another state, what further transitions can be made? Is it still possible to access all of the other configurations that are accessible from the polarised configuration? Can transitions be reversed? This is important for information storage applications, where it is important that the state of the system can be easily and reliably `re-written'.

\begin{figure}
  \centering
  \includegraphics[width=0.6\columnwidth]{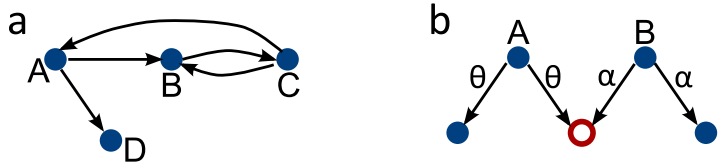}
  \caption{(a) An example directed network containing 4 nodes. The nodes $A$, $B$, $C$ form a strongly connected component, but $D$ is not a member of the component because there is no path from $D$ to the other nodes. (b) An example of how a node may be accessible, but not reproducibly so. The highlighted node can be reached by applying a field of angle $\theta$ to node $A$, or a field of angle $\alpha$ to node $B$, but both of these field angles also activate links to other nodes, so it is not possible to construct a field sequence that is guaranteed to pass through the highlighted node.}
  \label{small_network}
\end{figure}

A useful network theoretic concept here is that of strongly connected components (SCCs).
An SCC of a directed network is a set of nodes for which paths exist between every pair of nodes, taking the directions of the links into account. For example, in the network shown in figure~\ref{small_network}(a), the nodes $A$, $B$, $C$ form an SCC, but $D$ is not a member of the component because there is no path from $D$ to the other nodes. We determine the SCCs of a network using the algorithm in \cite{Tarjan:1972}, as implemented by the software package \textit{Mathematica}. 

Dynamics within an SCC are reversible, provided the correct field protocol is applied. In terms of networks, it is possible to travel along network links from any node of the SCC to any other node of the SCC; in terms of artificial spin ice dynamics this means that for any configuration in the SCC, there exists a sequence of fields to drive the system from that configuration to any other configuration in the SCC, and back again.

\begin{figure}
  \centering
  \includegraphics[width=\columnwidth]{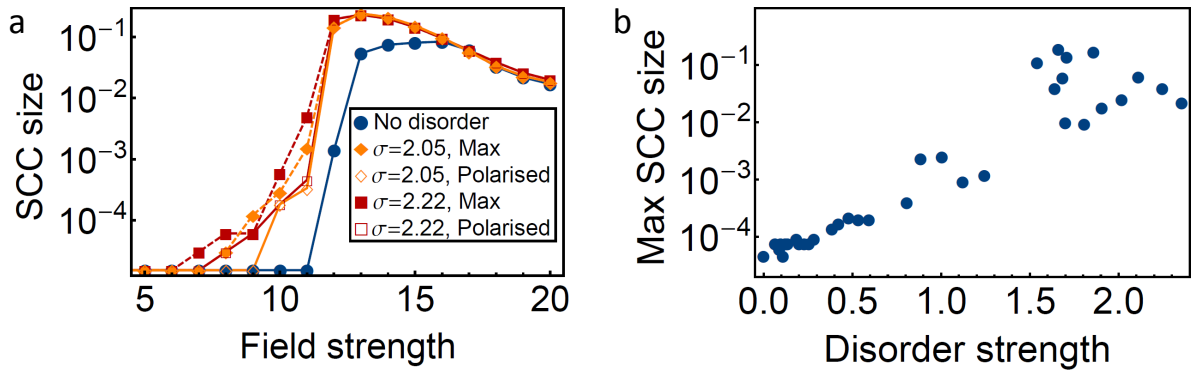}
  \caption{(a) Fraction of all nodes that are in the largest strongly connected component (SCC) and the SCC that contains the $+x$ polarised configuration, as a function of field strength. For the system without disorder, the $+x$ polarised configuration is always in the largest SCC. (b)Fraction of all nodes that are in the largest SCC, \textit{vs} disorder strength (standard deviation in island switching fields). The applied field strength is $h=11.5$. Each point represents a single disorder realisation.}
  \label{scc_vs_field_dis}
\end{figure}

The sizes of two SCCs in particular are of interest: the SCC that contains the $+x$ polarised configuration (or one of the other three polarised configurations: this choice is arbitrary), and the size of the largest SCC in the network. 
Figure~\ref{scc_vs_field_dis}(a) shows these sizes, plotted against field strength, for the system without disorder and the two disordered systems.  For the system without disorder, the $+x$ polarised configuration is always in the largest SCC.

Below a threshold field, all SCCs have size 1, that is, no dynamical transitions can be reversed. Above the threshold, the size of the largest SCC grows by three orders of magnitude to take in approximately ten percent of all nodes, before decreasing again for large field strengths. In the limit of very strong fields the largest SCC has size 4, and consists of the four polarised configurations. This limit holds for both perfect and disordered systems because strong external fields overcome disorder. When disorder is present, the $+x$ polarised configuration is not always in the largest SCC, but it is when the external field is sufficiently strong.

Comparison of figures \ref{n_reachable_polarised} and \ref{scc_vs_field_dis} indicates the strong correlation between the number of accessible states and SCC size. In general, the number of states accessible from a node must be greater than or equal to the size of its SCC. Figure~\ref{small_network}(a) illustrates a simple example of this: the number of nodes reachable from $A$ is $4$, while it is in an SCC of size 3. In fact, in the perfect system,  for $h\ge12$, all four polarised configurations are in the largest SCC and for $h\ge14$ all states that can be reached from the $+x$ polarised configuration are in the same SCC. Similar results hold for the disordered systems. 

The existence of large SCCs that include the polarised configurations implies that for correctly-tuned fields, the information storage capacity of the artificial spin ice is maximised, with several thousand configurations accessible from one another, making it possible to `write' a configuration and then `rewrite' a new configuration by applying a suitable sequence of fields. However, one should be careful for two reasons. First, if the number of states accessible from a given starting state (e.g, a polarised configuration) is larger than the SCC size, there will be dynamical `dead ends', that is, states that can be entered but not exited. Second, the existence of a path into a node does not guarantee that it is possible to reliably access that configuration. This is illustrated in figure~\ref{small_network}(b), where the highlighted node can be accessed, but it is not possible to construct a field sequence that is guaranteed to pass through it. A more detailed study of these two points is a topic for future work.

We close this section by commenting on the effect of disorder. While the general trends for SCC sizes as a function of field strength are the same for disordered and perfect systems,  for applied fields close to the mean island switching field of $11.25$, the difference in SCC size between perfect and disordered systems is substantial: around two orders of magnitude for $h=11$.  Figure~\ref{scc_vs_field_dis}(b) shows the size of the largest SCC for a range of disorder realisations, at an applied field strength of $h=11.5$. The size of the largest SCC increases continuously with disorder strength, although, as might be expected, the spread in values for different realisations of strong disorder is substantial.

Although the size of the largest SCC grows with disorder strength in the regime we study, we note that for a given array size and field strength, there is an upper bound on the number of accessible states, which is proportional to $2^{N'}$, where $N'$ is the number of spins that are not pinned and that do not always align with the external field, that is, the number of spins that are not `frozen' by disorder. This quantity decreases with disorder strength. This argument suggests there should be an optimal disorder strength for accessibility of states, but the optimal value depends on array size and field strength, rather than being universal to artificial spin ices.

\section{Network structure,`rewiring', and control}
\label{structure}

We saw in section~\ref{degrees} that the energetics of artificial spin ice plays a key role in determining network properties on a local, i.e: node, level. Similarly, in section \ref{accessibility}, we have seen how the global properties of the network relate to dynamics, which can be observed in simulation and are experimentally testable. In this section, we comment on the relationship between the local and global scales of the network, showing that purely local network properties are insufficient to predict the global structure and that instead, correlations exist in the network that are determined by the dynamics of artificial spin ice. We then show how the insight this gives about the network structure can be used to explore the possibility of controlling artificial spin ices.

We first demonstrate the existence of correlations in the artificial spin ice dynamics networks. We do this by comparing the spin ice networks with two other type of networks that retain some properties of the spin ice dynamics networks but are otherwise randomised.
Comparison with randomised networks is frequently used to reveal the structure of real-world networks, see, for example, \cite{Maslov2002, Guelzim2002, Eriksen2003, Kitsak2010a}.  

The first type of randomised network consists of uncorrelated random networks \cite{Erdos1959}, in which any link $i \to f$ exists with equal probability, $p=n(\mathrm{links})/2^{32}$ where $n(\mathrm{links})$ is the number of links and $2^{32}$ is the total number of possible links between $2^{16}$ nodes.   
The uncorrelated random networks represent a dynamics in which the possibility to pass from any state to any other state does not depend on energy at all. In particular, the relationship between the energy of a configuration and the degree of its network node is lost.

A more sophisticated approach to constructing random networks is to preserve this relationship, by creating `maximally random' networks consistent with a given joint in- and out-degree distribution~\cite{Newman2001, Milo2002} (see section~\ref{degrees}).
Under such a scheme, high-energy, unstable configurations have many links out, and low-energy, stable configurations have few links out, for example. This gives a network that, at least locally, represents a dynamics much more closely approximating the actual artificial spin ice dynamics.

\begin{figure}
  \centering
  \includegraphics[width=0.5\columnwidth]{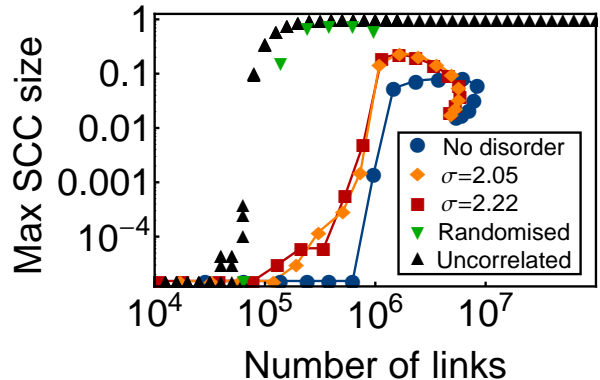}
  \caption{The fraction of nodes in the largest strongly connected component (SCC), \textit{vs} number of network links for spin ice dynamics networks and `test' networks with fewer correlations. In the `uncorrelated' test network, any link $i \to f$ exists with equal probability. In the `randomised' test networks, the joint in- and out-degree distributions are the same as those of the undisordered spin ice network,  but the links themselves are random.}
  \label{scc_vs_nedges}
\end{figure}

The comparison between these networks and the spin ice dynamics networks is illustrated in figure~\ref{scc_vs_nedges}, where we plot the fraction of nodes in the largest SCC (strongly connected component; see section~\ref{accessibility}) against the number of links in the network\footnote{A note on definitions: If nodes $A$ and $B$ are linked in both directions, that is, $(A\to B)$ and $(B \to A)$, we count each directed link separately, to give two links. Unless otherwise specified, we do not count self-links, that is links pointing from a node to itself. Under this definition, the number of links is given by $\sum_{i \ne j} \tilde{A}_{ij}$.}. For the spin ice dynamics networks, the size of the largest SCC is not a single-valued function of the number of links. However, as seen in figure~\ref{scc_vs_field_dis}(a), it is a single-valued function of field strength, which parameterises the curve, and the `doubling back' of  the SCC size \textit{vs} the number of links occurs because the number of links has a peak near $h=17$.

The randomised networks have very different global properties to the spin ice dynamics networks. This is because the connections in the spin ice dynamics networks are not simply dictated by the degree distribution, which in turn is because the dynamics are not simply dictated by the energies of states but also by the barriers between them.  While all networks show an increase in the size of the largest SCC with number of network links, the spin ice dynamics networks have a much reduced tendency towards large SCCs than the randomised networks do. SCC sizes are commonly used as a measure of percolation in directed networks \cite{Newman2001, Dorogovtsev2001, Schwartz2002, Boguna2005, AngelesSerrano2007,  Restrepo2008}, but, unlike the randomised networks, the spin ice dynamics networks never fully percolate. This may indicate a high level of clustering in the network \cite{Newman2003}.  

An intriguing feature of figure~\ref{scc_vs_nedges} is the dramatic increase in the size of the largest SCC in the artificial spin ice networks caused by a relatively small increase in the number of network links.
When the SCC is enlarged by tuning the field strength, the corresponding increase in the number of links is approximately ten-fold: from  $622,896$ to $6,180,266$ in the field window over which the largest SCC size in the perfect system increases from  1 node ($10^{-5}$ of all nodes) to 8\% of all nodes. When disorder is used to increase the SCC size, the change in the number of links is even smaller: at fixed field strength $h=11.5$, disorder increases the number of links from $736,720$ in a perfect system to $1,060,814$ when the disorder strength is $\sigma=3.3$ -- a change that is associated with the largest strongly connected component size growing from 3 nodes ($10^{-5}$ of all nodes) to 19\% of all nodes.

The fact that a relatively small change in the links of the network can alter the connectedness of nodes so dramatically suggests that for fields close to the mean switching field of $11.25$ the perfect system is `almost' well connected. This notion is supported by the jump in the number of states accessible from a polarised configuration at $h=11.5$ when disorder is turned on, shown in figure~\ref{n_reachable_polarised}(b), which we have already seen is caused by disorder allowing spins to flip that in the perfect system have a switching barrier slightly higher than the applied field.

The possibility for small changes to the system to have large effects on the accessibility of states finds application in the control of artificial spin ice. 
For example, in experimental studies of field-driven reversal in artificial kagome spin ice, islands were deliberately modified in order to serve as `start' and `stop' sites for avalanches of spin flips \cite{Mengotti:2010}, and simulations of a colloidal model for artificial spin ice reveal that using different barrier heights for different sublattices leads to a rich array of stable states that are different to those seen when all barriers are the same~\cite{Reichhardt2011}. As an alternative, one might imagine a system where a small number of macrospins are controlled directly via, for example, current-driving switching. In a network picture, such modifications of the spin ice system are equivalent to deliberately creating and removing certain links. As seen already in this section, such re-wiring can have a dramatic effect.

\begin{figure}
  \centering
  \includegraphics[width=0.5\columnwidth]{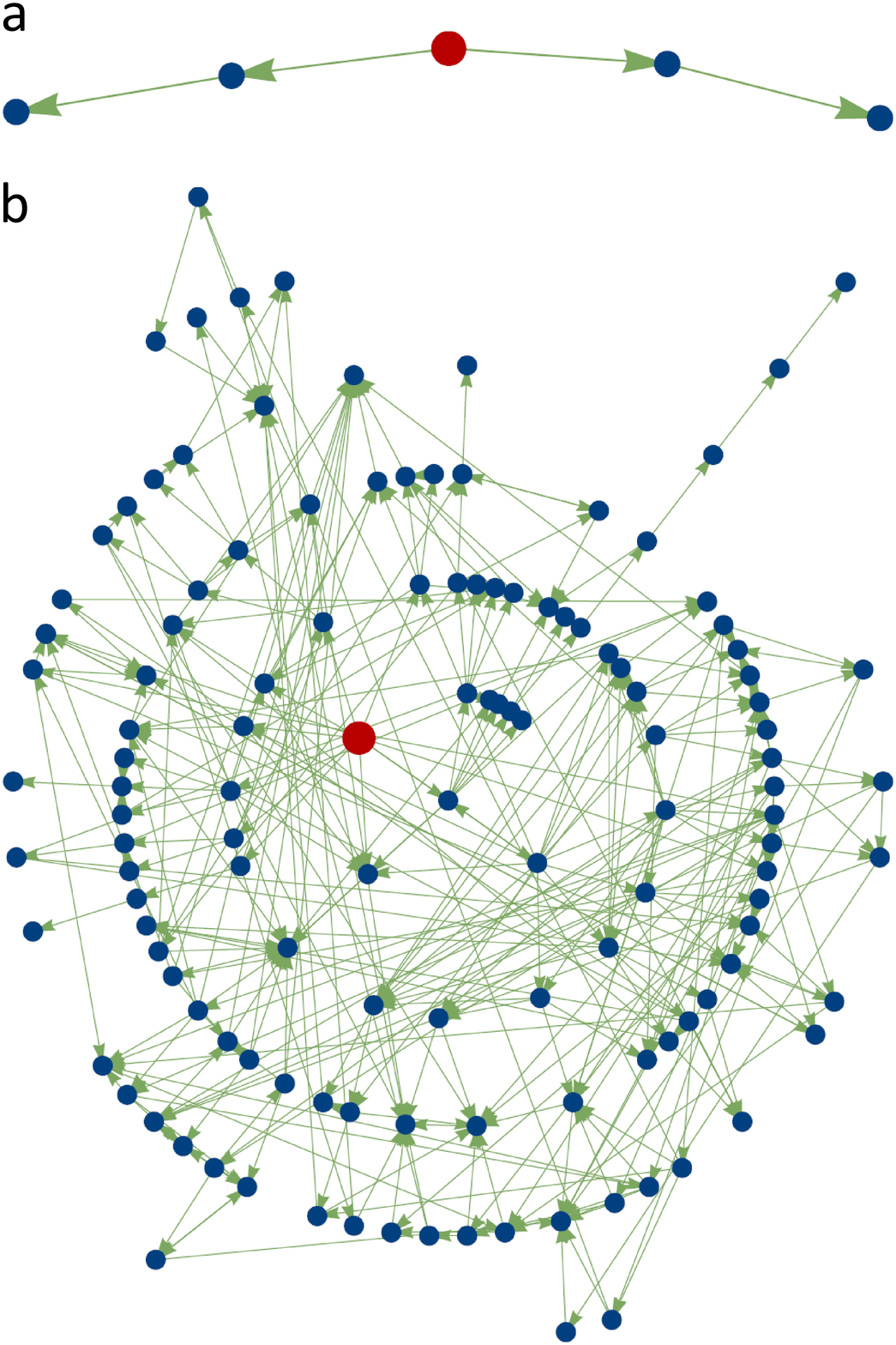}
  \caption{The network of states accessible from the $+x$ polarised configuration (the large red node), at a field strength of $h=11.5$, for (a) a perfect system, and (b) a system where the lower-left corner island can be flipped independently of the others.}
  \label{control}
\end{figure}

This effect is illustrated by the networks shown in figure~\ref{control}. Network (a) is the network of states accessible from the $+x$ polarised configuration (the large red node) for a perfect system, at a field strength of $h=11.5$. There are two other configurations accessible via a single field application, and a further two accessible if a second field is applied, making a total of 5 nodes. In contrast, network (b) is the network of configurations accessible from the same initial configuration, for the same field strength, but in a system where the lower-left corner island can be flipped independently of the others. This network contains 128 nodes. In other words, the ability to separately control a single spin yields an order-of-magnitude increase in the number of states accessible from the $+x$ polarised configuration.

These preliminary results demonstrate the value of the network picture of artificial spin ice dynamics for studying these problems. Future work might take advantage of other network properties. For example, in studies of how epidemics spread on networks, tools have been developed to determine which nodes are most important in determining the properties of transport on the network \cite{Kitsak2010a}. Since field-induced dynamics are essentially the same as network transport in this picture, applying these tools to networks describing spin ices may offer a way to determine how to modify the spin ice to control dynamics as desired.

\section{Conclusion and outlook}
\label{conclusion}
To summarise: we have shown that a network model for the dynamics of artificial spin ice provides a means of quantifying how applied field strength and quenched disorder affect the system's behaviour. Increasing disorder strength and tuning the applied field increase the number of states accessible to field-driven dynamics and the reversibility of dynamical transitions, both of which are important for potential applications. The changes in dynamics are caused by a `re-wiring' of the network that involves relatively few links, suggesting that the highly restricted dynamics of a perfect system subject to a sub-optimal field are caused by a small number of dynamical pathways being blocked. We have shown that, indeed, a small change to the artificial spin ice system unblocks these pathways and allows many new configurations to be accessed. We have also shown that the degree of a network node, a local property of the network, can also be related to the physics of the system, via the correlation between node degree and the energy of the configuration it represents.

It would be interesting to apply the network tools we have developed here to understand other problems in artificial spin ice. For example, the square ice studied here is only partially frustrated, and has a well-defined ground state even in the limit of short-range dipolar interactions. On the other hand, artificial kagome ice \cite{Tanaka:2006, Qi:2008} or a proposed square ice with sublattice height offsets \cite{Moller:2006, Mol:2010} are fully frustrated with extensive degeneracy, at least when interactions are short-ranged~\cite{Moller:2006, Moller2009, Chern2011}. Experimental studies have also been made comparing the demagnetisation of frustrated and unfrustrated small clusters of islands~\cite{Li:2010a}. Comparing the network properties of a fully frustrated ice with the partially frustrated square ice may give clues to the role of frustration in the dynamics.

There are also open avenues of enquiry related to the network structures themselves. For example, in studies of frustrated triangular antiferromagnets and the six vertex model, Peng \textit{et al.}~\cite{Peng2011} find fractal structures in the phase space networks which they suggest may be a signature of `long-range interactions, correlations, or boundary effects in real space'. The artificial spin ice systems studied here exhibit long range interactions, but whether their phase space networks have fractal structure is not yet known.

\ack{Z.B. thanks Paul Abbott for advice on the implementation of network construction algorithms. Z.B. and R.L.S. acknowledge the Australian Research Council and the Worldwide University Network for funding. Z.B. acknowledges funding from INFN and the Hackett Foundation.}

\end{document}